\documentclass[referee]{raa}
\usepackage{graphicx,times}
\usepackage{natbib}
\usepackage{amssymb,amsmath}
\usepackage{multirow}
\bibpunct{(}{)}{;}{a}{}{,}

\usepackage[a4paper=true,dvipdfm=true,pagebackref=true]{hyperref}
\hypersetup{pdftitle = The title of my PDF, pdfauthor = My name, pdfsubject= The subject, pdfkeywords = keyword1 keyword2 keyword3}
\hypersetup{colorlinks = true, linkcolor = green, anchorcolor = red, citecolor = blue, filecolor = red, pagecolor = red, urlcolor = red}

\begin{document}

   \title{Probing into the Possible Range of the U Bosonic Coupling Constants in Neutron Stars Containing Hyperons}

 \volnopage{ {\bf 2012} Vol.\ {\bf X} No. {\bf XX}, 000--000}
   \setcounter{page}{1}

   \author{Yan Xu\inst{1,2}, Bin Diao\inst{1,2}, Yi-Bo Wang\inst{1}, Xiu-Lin Huang\inst{1}, Xing-Xing Hu\inst{1}, Zi Yu\inst{3}}

   \institute{ Changchun Observatory, National Astronomical Observatories,
Chinese Academy of Sciences, Changchun 130117, China; {\it Corresponding Author. Y.Xu, xuy@cho.ac.cn; Z.Yu, ziyu$_{-}$njfu@163.com}\\
   \and School of Astronomy and Space Sciences, University of Chinese Academy of Sciences, Beijing 100049, China\\
   \and College of Science, Nanjing Forestry University, Nanjing 210037, China\\
\vs \no
   {\small Received xxxx June xx; accepted xxxx July xx}
}

\abstract{The range of the U bosonic coupling constants in neutron star matter is a very interesting but still unsolved problem which has multifaceted influences in nuclear physics, particle physics, astrophysics and cosmology.
The combination of the theoretical numerical simulation and the recent observations provides a very good opportunity to solve this problem. In the present work, the range of the U bosonic coupling constants is inferred based on the three relations of the mass-radius, mass-frequency and mass-tidal deformability in neutron star containing hyperons using the GM1, TM1 and NL3 parameter sets under the two flavor symmetries of the SU(6) and SU(3) in the framework of the relativistic mean field theory. Combined with observations from PSRs J1614-2230, J0348+0432, J2215-5135, J0952-0607, J0740+6620, J0030-0451, J1748-2446ad, XTE J1739-285, GW170817 and GW190814 events, our numerical results show that the U bosonic coupling constants may tend to be within the range from 0 to 20 GeV$^{-2}$ in neutron star containing hyperons. Moreover, the numerical results of the three relations obtained by the SU(3) symmetry are better in accordance with observation data than those obtained by the SU(6) symmetry. The results will help us to improve the strict constraints of the equation of state for neutron stars containing hyperons. }

   \authorrunning{Y. Xu et al. }            
   \titlerunning{the U Bosonic Coupling Constants in Neutron Stars}  
   \maketitle

%
\section{Introduction}           
\label{sect:intro}
In May 1932, Chadwick discovered that there is a new uncharged particle in the nucleus, named neutron(\citealt{Landau+1932}).
It made people realize that the nucleus is composed of neutrons and protons.
In the same year, Landau predicted that the possible existence of stars which is composed of neutrons, who called them "gigantic nuclei".
In 1934, Baade and Zwicky separately proposed that neutron stars are born in supernova explosions, who coined the term "neutron stars"(\citealt{Baade+Zwicky+1932}).
In 1939, Oppenheimer and Volkoff first performed the theoretical calculations on the structure of neutron stars by using the work of Tolman on the spherical symmetry metric, who regarded as neutrons as an ideal degenerate Fermi gas and established the Tolman-Oppenheimer-Volkoff equations. This is the first quantitative neutron star model for deriving the mass limit of neutron star.
They found such stars would have a maximum mass of 0.7$M_{\odot}$ which is significantly lower than the maximum mass limit 1.4$M_{\odot}$ of white dwarf(\citealt{Oppenheimer+Volkoff+1939,Tolman+1939,Dar+2014}).
Up to now, the researches in the structures of neutron stars are more and more deeply with the continuous development in pulsar observations and the theoretical studies of neutron stars(\citealt{Han+2021,Pant+2021,Zhou+2022}).
Although many assumptions have been considered to calculate the mass limit and explored the internal composition of neutron stars, the above problems still remains mysteries due to the likely more complex structures of neutron stars.

In 2015, the gravitational waves(GWs) from the merger of a binary black hole were first directly detected, namely GW150914(\citealt{Abbott+etal+2016}).
After two years, the gravitational waves from the merger of a binary neutron star was discovered by the LIGO and Virgo, namely GW170817(\citealt{Abbott+etal+2017}).
It is the first time in human history that the gravitational wave observatory and the electromagnetic wave telescope are used to observe the same astrophysical event, marking the beginning of a multi-messenger era of neutron star research. In 2019, the gravitational waves from the merger of a binary compact star was discovered by the LIGO and Virgo, namely GW190814(\citealt{Abbott+etal+2020}). In particular, the recent astrophysical constraints on tidal deformability $\Lambda_{1.4}$ of a canonical 1.4 $M_{\odot}$ neutron star from the GW170817 and GW190814 are $\Lambda_{1.4}=190^{+390}_{-120}$ and $\Lambda_{1.4}=616^{+273}_{-158}$, respectively(\citealt{Abbott+etal+2018, Abbott+etal+2020}). The observed massive neutron stars mainly include PSRs J1614-2230 ($1.908\pm0.016$ $M_{\odot}$) (\citealt{Arzoumanian+etal+2018}), J0348+0432 ($2.01\pm0.04$ $M_{\odot}$) (\citealt{Antoniadis+etal+2013}), J0740+6620 ($2.072^{+0.067}_{-0.066}$ $M_{\odot}$ with radius $12.39^{+1.30}_{-0.98}$ km) (\citealt{Cromartie+etal+2020,Fonseca+etal+2021,Riley+etal+2021,Ferreira+etal+2021}) and J2215+5135 ($2.27^{+0.17}_{-0.15}$ $M_{\odot}$)(\citealt{Linares+etal+2018}), etc. Especially PSR J0952-0607 with $2.35\pm0.17$ $M_{\odot}$ which is already far more than the masses of the previous massive neutron stars(\citealt{Kuzuhara+etal+2022,Ecker+etal+2023}). In addition, the mass and radius of PSR J0030-0451 have estimated by NICER mission, respectively($1.34^{+0.15}_{-0.16}$ $M_{\odot}$, $12.71^{+1.14}_{-1.19}$ km and $1.44^{+0.15}_{-0.14}$ $M_{\odot}$, $13.02^{+1.24}_{-1.06}$ km)(\citealt{Riley+etal+2019,Li+etal+2021}).
In all, the breakthrough detections of the GW170817 and GW190814 by the LIGO and Virgo, and the precise measurements such as the mass, radius and rotation frequency of neutron star by multi-messenger advanced telescopes such as FAST, NICER, and SKA have jointly provided a large amount of reliable observational data for the research of neutron star structure and internal composition(\citealt{Abbott+etal+2018,Riley+etal+2019,Miller+etal+2019,Chatziioannou+2020,Li+etal+2021,Gao+etal+2022}). However, many of us face the questions: whether there are new
degrees of freedom such as hyperons besides nucleons in neutron stars? What the kind of interactions would be allowed the appearance of the new degrees of freedom in the massive neutron stars(\citealt{Xia+etal+2014,paper+2019,Zhao+2020,Khadkikar+etal+2022,Zhao+Liu+2022,Sun+etal+2023})?

In 1980, Fayet firstly proposed that the possible existence of a neutral weakly coupling light spin-1 vector U boson which has been considered to provide a new kind of repulsive interaction in neutron stars(\citealt{Fayet+1980}). The researches shows that the interaction between U bosons and baryons can modify the equation of states, gravities and increase the maximum masses of neutron stars obviously (\citealt{Krivoruchenko+etal+2009,Yu+etal+2011,Zhang+etal+2011,Zhang+etal+2016,Ding+etal+2016,Yu+etal+2018}). However, there is little work about whether and how U bosons affect the tidal deformation properties of neutron stars. In addition, though the range of the U bosonic coupling constants in neutron star matter has aroused a wide attention, it remains unsolved. The comparison between the theoretical results of neutron stars and the observed data provides a possible solution to solve these problems.

In the work, the equation of states, the three relations of the mass-radius, mass-frequency and mass-tidal deformability of neutron stars containing hyperons are obtained using the GM1, TM1 and NL3 parameter sets in the relativistic mean field theory including the U bosons(\citealt{Glendenning+1985,Bednarek+Manka+2005,Yang+Shen+2008,Miyatsu+etal+2013}). A combination of the observed data from PSRs J1614-2230, J0348+0432, J2215-5135, J0952-0607, J0740+6620, J0030-0451, J1748-2446ad, XTE J1739-285, GW170817 and GW190814 events, we will mainly explore the possible range of the U bosonic coupling constants based on the macroscopic properties of neutron stars containing hyperons such as mass, radius, frequency and tidal deformability.
\section{Theoretical framework}
\subsection{Relativistic Mean Field Theory}
\label{sect:Obs}
Here we apply the relativistic mean field approximation including $\sigma$, $\omega$, $\rho$, $\sigma^{*}$ and $\phi$ mesons to describe the properties of the cores of neutron stars(\citealt{Miyatsu+etal+2013,Xu+etal+2015,Xu+etal+2021,Miyatsu+etal+2022}). The total Lagrangian density can be expressed as
\begin{eqnarray}
\label{eq:L}
\mathcal{L}=\sum_B\mathcal{L}_{B}+\sum_l\mathcal{L}_{l}+\mathcal{L}_{m}+\mathcal{L}_{u},
\end{eqnarray}
where B, l and m represents the baryons, leptons and mesons, respectively. The sum on B runs over the octet baryons(n, p, $\Lambda$, $\Sigma^{+}$, $\Sigma^{0}$, $\Sigma^{-}$, $\Xi^{-}$ and $\Xi^{0}$), and the sum on l is over electrons and muons(e and $\mu$). Explicitly, the Lagrangian is
\begin{eqnarray}
\mathcal{L}_{B}=\overline{\psi}_B[i\gamma_\mu\partial^\mu-(m_B-g_{\sigma B}\sigma-g_{\sigma^*B}\sigma^*)-g_{\rho B}\gamma_{\mu}{\vec{\mathbf{\rho}}^\mu}\cdot\vec{I}_{B}-g_{\omega B}\gamma_\mu\omega^\mu-g_{\phi B}\gamma_\mu\phi^\mu]\psi_B,\\
\mathcal{L}_{l}=\overline{\psi}_l[i\gamma_\mu\partial^\mu-m_l]\psi_l,\\
\mathcal{L}_{m}=-\frac{1}{4}W^{\mu v}W_{\mu v}-\frac{1}{4}\vec{R}^{\mu v}\cdot\vec{R}_{\mu v}-\frac{1}{4}P^{\mu v}P_{\mu v} +\frac{1}{2}m_\omega^2\omega_\mu\omega^\mu+\frac{1}{2}m_\rho^2{\vec{\mathbf{\rho}}}_\mu\cdot{\vec{\mathbf{\rho}}}^\mu+\frac{1}{2}m^2_{\phi}\phi_\mu\phi^\mu \nonumber \\
+\frac{1}{2}(\partial_\mu\sigma\partial^\mu\sigma-m_\sigma^2\sigma^2)+\frac{1}{2}(\partial_v\sigma^*\partial^v\sigma^*-m^2_{\sigma^*}\sigma^{*2})-\frac{1}{3}a\sigma^{3}-\frac{1}{4}b\sigma^4+\frac{1}{4}c_3(\omega_\mu\omega^\mu)^2,
\end{eqnarray}
Moreover, we consider the contribution of the U bosons to the total Lagrangian density. On the basis of the conventional view, the Yukawa term correction to Newtonian gravity is only applied at the matter part without the
geometric part in general relativity(\citealt{Fujii+1971,Fujii+1988}). $\mathcal{L}_{u}$ is given by
\begin{eqnarray}
\mathcal{L}_{u}=-\overline{\psi}_Bg_{uB}\gamma_\mu u^\mu\psi_B-\frac{1}{4}U^{\mu v}U_{\mu v}+\frac{1}{2}m_u^2u_\mu u^\mu.
\end{eqnarray}
Here $\psi_{B}$ and $\psi_{l}$ are the baryonic and leptonic Dirac fields, respectively. $\gamma_{u}$ represents the Dirac matrice. $\vec{I}_{B}$ is the baryonic isospin matrix. $g_{mB}$ ($g_{uB}$) denotes the mesonic (U bosonic) and baryonic coupling constants. The field strength tensors of $\omega$, $\rho$, $\phi$ and U can be defined as $W_{\mu v}=\partial_\mu\omega_v-\partial_v\omega_\mu$, $R_{\mu v}=\partial_\mu{\mathbf{\rho}}_v-\partial_v{\mathbf{\rho}}_\mu$, $P_{\mu v}=\partial_\mu\phi_v-\partial_v\phi_\mu$ and $U_{\mu v}=\partial_\mu u_v-\partial_v u_\mu$, respectively. $m_B$, $m_l$ and $m_u$ are the masses of baryons, leptons and U bosons, respectively.

According to the standard Euler-Lagrange equation, the equations of motion of baryons, mesons and U bosons can be derived from the total Lagrangian density in Eq.(1). The fields of the five mesons and U bosons are replaced by their classical expectation values in the relativistic mean field approximation, thus the equations of motion are given as follows
\begin{eqnarray}
\label{eq: baryon and meson fields}
(i\gamma_{\mu}\partial^{\mu}-m_B+g_{\sigma B}\sigma&+&g_{\sigma^*B}\sigma^*-g_{\omega B}\gamma_{0}\omega^{0}-g_{\rho B}\gamma_{0}I_{3B}\rho^{0}_{3}-g_{\phi B}\gamma_{0}\phi^0-g_{uB}\gamma_0 u^0)\psi_{B}=0,\\
& &\sum_B g_{\sigma B}n_{SB}=m_\sigma^2\sigma+a\sigma^2+b\sigma^3,\\
& &\sum_B g_{\omega B}n_B=m_\omega^2\omega_0+c_{3}\omega^{3}_{0},\\
& &\sum_B g_{\rho B}n_{B}I_{3B}=m_{\rho}^2\rho_{03},\\
& &\sum_B g_{\sigma^* B}n_{SB}=m_{\sigma^*}^2\sigma^*,\\
& &\sum_B g_{\phi B}n_{B}=m_\phi^2\phi_0,\\
& &\sum_B g_{u B}n_{B}=m_u^2 u_0.
\end{eqnarray}
Here, $m_B^{*}=m_B-g_{\sigma B}\sigma-g_{\sigma^*B}\sigma^*$ is the baryonic effective mass. $n_{SB}$ and $n_{B}$ represent the scalar and baryonic number densities, respectively. The equations of state within the framework of the relativistic mean field theory can be written as
\begin{eqnarray}
\label{eq:the equations of state}
\varepsilon=\varepsilon_{0}+\varepsilon_{uB},\ \ \ P=P_{0}+P_{uB}.
\end{eqnarray}
Here $\varepsilon_{0}$ and $P_{0}$ are the the conventional energy density and pressure in the neutron star matter, respectively. Their formulas as follows:
\begin{eqnarray}
\label{eq: energy and press}
\varepsilon_{0}&=&\sum_B\frac{1}{\pi^{2}}\int_0^{p_{FB}}\sqrt{p_{B}^{2}+m_{B}^{*2}}p_{B}^{2}dp_{B}+\frac{1}{2}m_{\sigma}^{2}\sigma^{2}+\frac{1}{3}a\sigma^{3}\nonumber \\
&+&\frac{1}{4}b\sigma^4+\frac{1}{2}m_{\sigma^{*}}^{2}\sigma^{*2}+\frac{1}{2}m_{\omega}^{2}\omega^{2}+\frac{3}{4}c_{3}\omega^{4}+\frac{1}{2}m_{\phi}^{2}\phi^{2} \nonumber \\
&+&\frac{1}{2}m_{\rho}^{2}\rho^{2}+\sum_l\frac{1}{\pi^{2}}\int_0^{p_{Fl}}\sqrt{p_{l}^{2}+m_{l}^{*2}}p_{l}^{2}dp_{l},\\
P_{0}&=&\frac{1}{3}\sum_B\frac{1}{\pi^{2}}\int_0^{p_{FB}}\frac{p_{B}^{4}dp_{B}}{\sqrt{p_{B}^{2}+m_{B}^{*2}}}-\frac{1}{2}m_{\sigma}^{2}\sigma^{2}-\frac{1}{3}a\sigma^{3}\nonumber \\
&-&\frac{1}{4}b\sigma^4-\frac{1}{2}m_{\sigma^{*}}^{2}\sigma^{*2}+\frac{1}{2}m_{\omega}^{2}\omega^{2}+\frac{1}{4}c_{3}\omega^{4}+\frac{1}{2}m_{\phi}^{2}\phi^{2}\nonumber\\
&+&\frac{1}{2}m_{\rho}^{2}\rho^{2}+\frac{1}{3}\sum_l\frac{1}{\pi^{2}}\int_0^{p_{Fl}}\frac{p_{l}^{4}dk}{\sqrt{p_{l}^{2}+m_{l}^{*2}}p_{l}^{2}}dp_{l},
\end{eqnarray}
where $p_{FB}$ expresses the baryonic Fermi momentum. Moreover, the bosonic additional contribution for the equations of state can be expressed as
\begin{eqnarray}
\label{eq:the equations of state}
\varepsilon_{uB}=P_{uB}=\frac{g_{uB}^2n_B^2}{2m_u^2}.
\end{eqnarray}
For the convenience of discussion, the baryonic and U bosonic coupling constant is defined as $g_{u}=(g_{uB}/m_u)^2$.
We can calculate the equations of state of neutron star matter under the $\beta$-equilibrium and charge neutrality conditions,
\begin{eqnarray}
\label{eq: chemical potential}
\mu_{B}=\mu_{n}-q_{B}\mu_{e},\ \ \ \sum_B q_{B}\rho_{B}=0.
\end{eqnarray}
Here $q_{B}$ expresses the baryonic electric charge in unit of e.
\subsection{Tolman-Oppenheimer-Volkoff equations}
The mass-radius relations of neutron stars can be obtained by inputing the numerical results of the equations of state in the Tolman-Oppenheimer-Volkoff equations(\citealt{Oppenheimer+Volkoff+1939,Tolman+1939})
\begin{eqnarray}
\label{eq: TOV}
\frac{dP(r)}{dr}&=&-\frac{[P(r)+\varepsilon(r)][M(r)+4\pi r^{3}P(r)]}{r(r-2M(r))},\\
\frac{dM(r)}{dr}&=&4\pi r^{2}\varepsilon(r).\\\nonumber
\end{eqnarray}
Here Eqs. (18) and (19) satisfy the boundary conditions $P(r=0)=0$, $P(r=R)=0$ and $M(r=0)=0$. $R$ is the radius of a neutron star.
\subsection{Tidal Deformability}
The dimensionless tidal deformability $\Lambda$ is defined as
\begin{eqnarray}
\Lambda = \frac{2}{3}k_2(M/R)^{-5}.
\end{eqnarray}
Here $k_2$ is the tidal Love number and describes the reaction of a neutron star to the appearance of the external tidal field. It is given by
\begin{eqnarray}
k_2 & = & \frac{1}{20}\biggl( \frac{2M}{R}\biggr)^5\biggl( 1-\frac{2M}{R}\biggr)^2\biggl[ 2-y_R + (y_R - 1)\frac{2M}{R}\biggr] \nonumber \\
& & \times \biggl\{\frac{2M}{R}\biggl(6 - 3y_R + \frac{3M}{R}(5y_R - 8) + \frac{1}{4}\biggl(\frac{2M}{R}\biggr)^2 \nonumber \\
& & \times\biggl[26 - 22y_R + \biggl(\frac{2M}{R}\biggr)(3y_R - 2) + \biggl(\frac{2M}{R}\biggr)^2(1+y_R)\biggr]\biggr) \nonumber \\
& & + 3\biggl(1-\frac{2M}{R}\biggr)^2\bigg[2 - y_R + (y_R-1)\frac{2M}{R} \bigg] \nonumber \\
& & \times \ln\biggl(1-\frac{2M}{R} \biggr) \biggr\}^{-1}.
\end{eqnarray}
The quantity $y_R$ can be obtained by solving the following differential equation
\begin{eqnarray}
r\frac{dy(r)}{dr} + y(r)^2 + y(r)F(r) +r^2Q(r) = 0.
\end{eqnarray}
Here Eq. (22) satisfies the initial condition $y(0)=2$. $F(r)$ and $Q(r)$ are
\begin{eqnarray}
F(r)&=&\frac{r-4\pi r^3[\mathcal{E}(r)-P(r)] }{r-2M(r)},\\
Q(r)&=&\frac{4\pi r\biggl( 5\mathcal{E}(r)+9P(r)+\frac{\mathcal{E}(r)+P(r)}{\partial P/\partial\mathcal{E}}-\frac{6}{4\pi r^2}\biggr)}{r-2M(r)}-4\biggl[\frac{M(r)+4\pi r^3P(r)}{r^2(1-2M(r)/r)}\biggr]^2.
\end{eqnarray}
It can be seen from these Eqs. (21-24) that the dimensionless tidal deformability $\Lambda$ is depending on the equations of state, the compactness $M/R$ and the quantity $y_R$ of a neutron star.
\begin{table}[]
\centering
\caption{\label{tab:parameter}
The three parameter sets(\citealt{Miyatsu+etal+2013}). $m_{\sigma}$ = 550, 508.194, 511.198 MeV, $m_{\omega}$ = 783, 783 and 782.501 MeV, $m_{\rho}$ = 770, 770 and 763 MeV and $m_{N}$ = 939, 938 and 939 MeV in GM1,TM1 and NL3, respectively(\citealt{Sugahara+1994,Lalazissis+etal+1997,Bank+2001,Lopes+2012}). The coupling relations are set to $g_{\sigma^{*}N}=g_{\rho_{\Lambda}}=0$ and $g_{\sigma^{*}\Lambda}=g_{\sigma^{*}\Sigma}=0$.
}
{\begin{tabular}{@{}ccccccccccccc@{}}
\hline
Parameter sets             & \multicolumn{2}{c}{GM1}   & \multicolumn{2}{c}{TM1}   & \multicolumn{2}{c}{NL3} \\
\hline
Symmetries                 & SU(6)      & SU(3)        & SU(6)       & SU(3)       & SU(6)        & SU(3)    \\
\hline
$g_{\sigma N}$             & 9.57   & 9.57           & 10.029     & 10.029     & 10.217  & 10.217        \\
$g_{2}$ (fm$^{-1}$)        & 12.28  & 12.28          & 7.233      & 7.233      & 10.431  & 10.431        \\
$g_{3}$                    & -8.98  & -8.98          & 0.618      & 0.618      & -28.885 & -28.885       \\
$c_{3}$                    & ---    & ---            & 71.308     & 81.601     & ---     & ---           \\
$g_{\omega N}$             & 10.61  & 10.26          & 12.614     & 12.199     & 12.868  & 12.450        \\
$g_{\rho N}$               & 4.10   & 4.10           & 4.632      & 4.640      & 4.474   & 4.474         \\
$g_{\phi N}$               & ---    & -3.50          & ---        & -4.164     & ---     & -4.250        \\
$g_{\sigma\Lambda}$        & 5.84   & 7.25           & 6.170      & 7.733      & 6.269   & 7.853         \\
$g_{\sigma\Sigma}$         & 3.87   & 5.28           & 4.472      & 6.035      & 4.709   & 6.293         \\
$g_{\sigma\Xi}$            & 3.06   & 5.87           & 3.202      & 6.328      & 3.242   & 6.408         \\
$g_{\sigma^{\ast}\Lambda}$ & 3.73   & 2.60           & 5.015      & 3.691      & 5.374   & 4.174         \\
$g_{\sigma^{\ast}\Xi}$     & 9.67   & 6.82           & 11.516     & 8.100      & 11.765  & 8.378         \\
\hline
\multicolumn{7}{c}{Properties of the symmetric nuclear matter}                                           \\
\hline
$n_{B}^{0}$ (fm$^{-3}$)    & 0.153  & 0.153          & 0.145      & 0.145      & 0.148   & 0.148         \\
$\omega_{0}$ (MeV)         & -16.3  & -16.3          & -16.3      & -16.3      & -16.299 & -16.299       \\
$m_{B}^{*}$$/m_{B}$        & 0.70   & 0.70           & 0.634      & 0.634      & 0.60    & 0.60          \\
$K_{\nu}$ (MeV)            & 300    & 300            & 281        & 284        & 271.76  & 271.76        \\
$a_{4}$ (MeV)              & 32.5   & 32.5           & 36.9       & 36.9       & 37.4    & 37.4          \\
$L$ (MeV)                  & 93.9   & 93.9           & 110.9      & 110.8      & 118.0   & 118.0         \\
\hline
\end{tabular} \label{tap}}
\end{table}
\subsection{Coupling constants}
To gain the numerical results of the equations of state and the properties of neutron star matter, we need three kinds of coupling constants in order to solving the above equations. The first one is the nucleonic and mesonic coupling constants which could reflect the saturation properties of the symmetric nuclear matter. The second one is the hyperonic and mesonic coupling constants. They are impossible to be determined by the saturation properties of the symmetric nuclear matter because the hyperonic critical densities are much higher than the saturation density of nuclear matter. The hyperonic and mesonic coupling constants can be determined so as to reproduce the observed properties of nuclear matter and the experimental data of hypernuclei. Thus the baryonic and mesonic coupling constants under SU(6) spin-flavor and SU(3) flavor symmetries are widely used in the study of neutron stars.  In two symmetries, the $\omega$ and $\phi$ mesons are described by the pure singlet ${\left|1\right\rangle}$ and octet ${\left|8\right\rangle}$, then read
\begin{eqnarray}
\omega=\cos\theta_{v}+\sin\theta_{v},\ \ \ \phi=-\sin\theta_{v}+\cos\theta_{v}.
\end{eqnarray}

In the SU(6) spin-flavor symmetry, the mixing angle and the coupling ratio are given by
\begin{eqnarray}
\theta_{v}^{ideal}=\simeq 35.26^{\circ},
\ \ \ z=\frac{1}{\sqrt{6}}\simeq 0.4082.
\end{eqnarray}
The usual SU(6) relations are as follows:
\begin{eqnarray}
g_{\omega N}=\frac{3}{2}g_{\omega\Lambda}=\frac{3}{2}g_{\omega\Sigma}=3g_{\omega\Xi},\ \ \ g_{\phi N}=0,\ \ \ g_{\phi\Lambda}=g_{\phi\Sigma}=\frac{1}{2}g_{\phi\Xi}=\frac{\sqrt{2}}{3}g_{\omega N}.
\end{eqnarray}

The SU(3) symmetry can be seen as the symmetry group with three flavors of up, down and strange quarks denoting the strong interactions. In the SU(3) flavor symmetry, the mixing angle and the coupling ratio are given by
\begin{eqnarray}
\theta_{v}^{ideal}=\simeq 37.50^{\circ},
\ \ \ z=0.1949.
\end{eqnarray}
The SU(3) relations are as follows:
\begin{eqnarray}
&&g_{\omega\Lambda}=g_{\omega\Sigma}=\frac{1}{1+\sqrt{3}z\tan\theta_{v}}g_{\omega N},\ \ \ g_{\omega\Xi}=\frac{1-\sqrt{3}z\tan\theta_{v}}{1+\sqrt{3}z\tan\theta_{v}}g_{\omega N}, \nonumber\\
&&g_{\phi\Lambda}=g_{\phi\Sigma}=\frac{-\tan\theta_{v}}{1+\sqrt{3}z\tan\theta_{v}} g_{\omega N},\ \ \ g_{\phi\Xi}=-\frac{\sqrt{3}z+\tan\theta_{v}}{1+\sqrt{3}z\tan\theta_{v}} g_{\omega N},\nonumber\\
&&g_{\phi N}=\frac{\sqrt{3}z-\tan\theta_{v}}{1+\sqrt{3}z\tan\theta_{v}} g_{\omega N}.
\end{eqnarray}
The third one is the baryonic and U bosonic coupling constants. In order to ensure the stability of neutron stars using the equations of state constrained by the
FOPI/GSI data, Wen et al. point out that $g_{u}$ should be located in 0-100 GeV$^{-2}$(\citealt{Xiao+etal+2009,Wen+etal+2009,Zhang+etal+2011}). In this work, the U bosonic influence is researched within the region of $g_{u}=$0-40 GeV$^{-2}$. Three typical parameter sets GM1, TM1 and NL3 are selected to describe the interactions and the corresponding the saturation properties are listed in Table 1.
\begin{figure}
   \centering
   \includegraphics[width=12.0cm, angle=0]{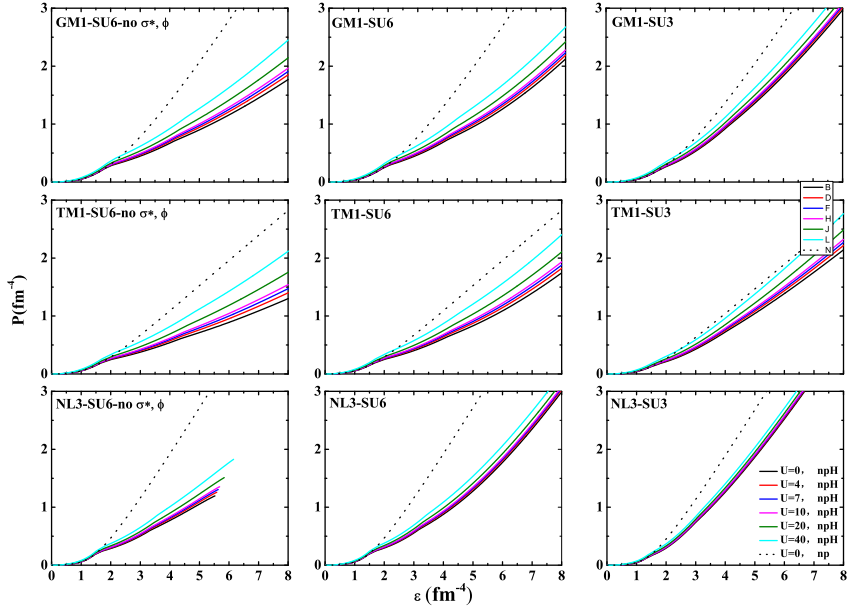}
   \caption{Pressures $P$ vs energy densities $\varepsilon$ of neutron stars for the three cases, SU6-no $\sigma^{*}$, $\phi$, SU6 and SU3 under each parameter set. In the paper, the the black, red, blue, magenta, olive and oyan colored lines express the six cases of $g_{u}$ = = 0, 4, 7, 10, 20 and 40 GeV$^{-2}$ in npH matter, respectively. The dot line stands for the case of $g_{u}$ = 0 GeV$^{-2}$ in np matter. }
   \label{Fig1}
   \end{figure}

\begin{figure}
   \centering
  \includegraphics[width=12.0cm, angle=0]{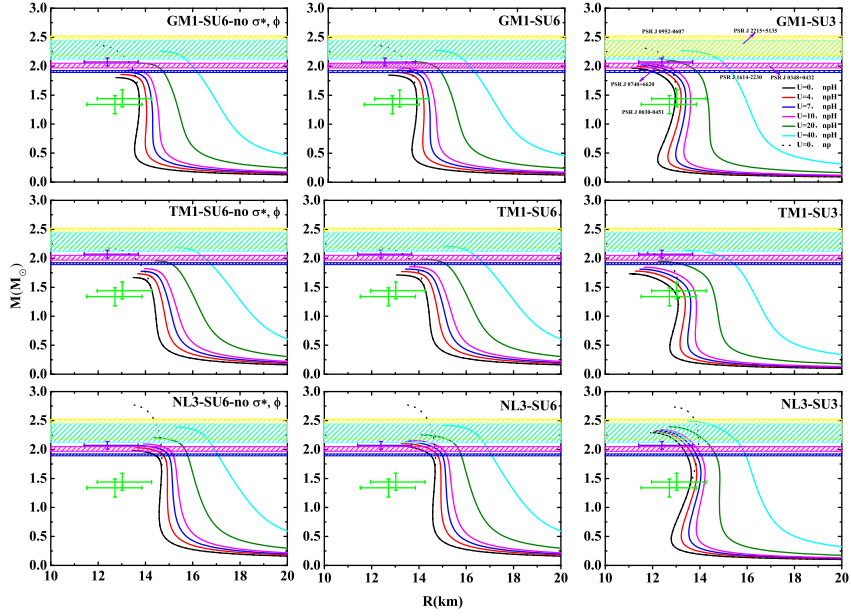}
   \caption{Masses vs radii of neutron stars for the three cases, SU6-no $\sigma^{*}$, $\phi$, SU6 and SU3 under each parameter set. Also shown are the recent constraints on the mass and radius from NICER measurements of PSRs J1614-2230, J0348+0432, J2215-5135, J0952-0607, J0030-0451 and J0740+6620.}
   \label{Fig2}
   \end{figure}

\begin{figure}
   \centering
  \includegraphics[width=12.0cm, angle=0]{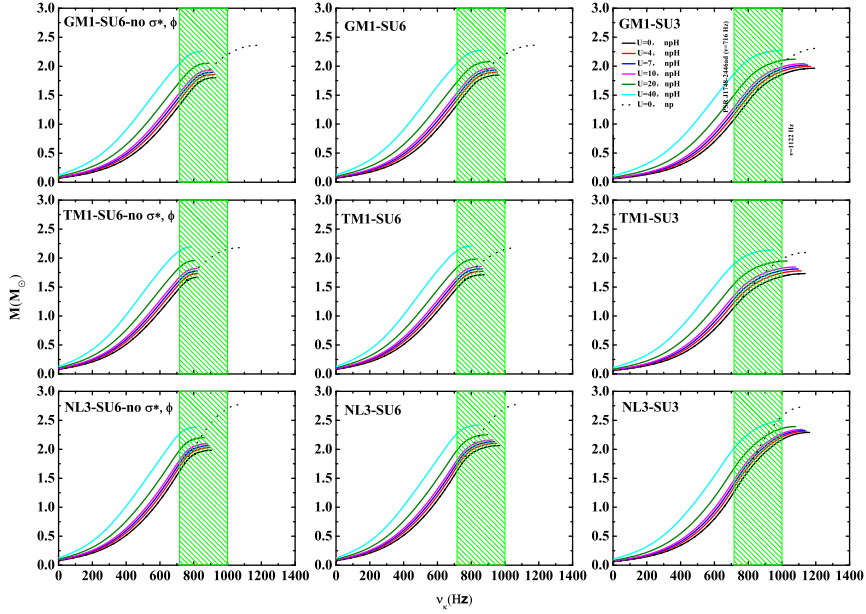}
   \caption{Masses vs Kepler frequencies of neutron stars for the three cases, SU6-no $\sigma^{*}$, $\phi$, SU6 and SU3 under each parameter set. The green shaded region corresponds to the observational limit on the frequency from the rapidly rotating pulsars PSR J1748-2446ad ($\nu$ = 716 Hz) and $\nu$ = 1000 Hz as an example to limit our results.}
   \label{Fig3}
   \end{figure}

\begin{figure}
   \centering
  \includegraphics[width=12.0cm, angle=0]{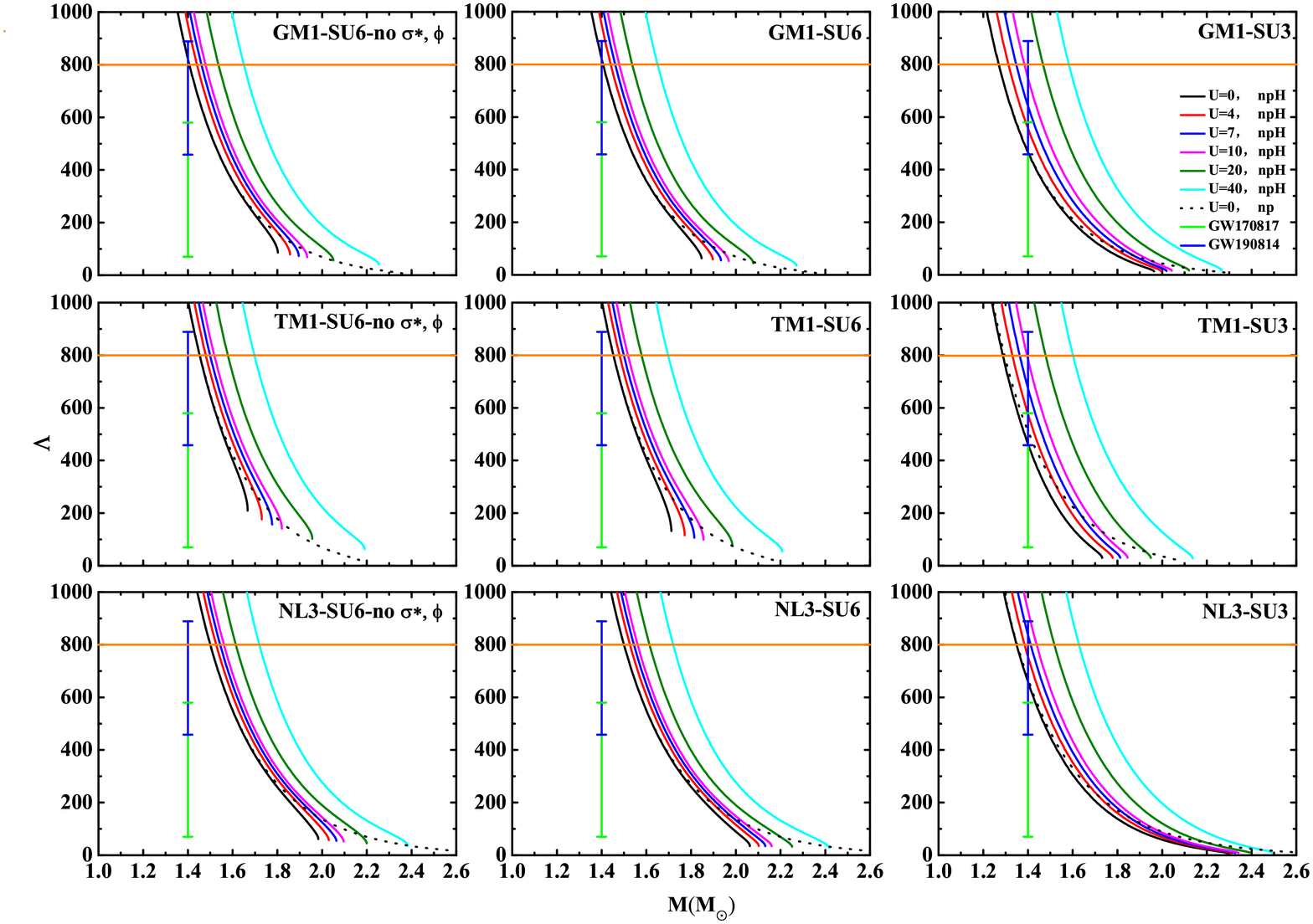}
   \caption{Tidal deformabilities vs masses of neutron stars for the three cases, SU6-no $\sigma^{*}$, $\phi$, SU6 and SU3 under each parameter set. The green and blue error bars represent the constraints on the tidal deformabilites of neutron stars by GW170817 and GW190814 events. The orange line corresponds to the up limit value of the $\Lambda_{1.4}$ from the GW170817 data.}
   \label{Fig4}
   \end{figure}

\section{Result and Discussion}
\label{sect:discussion}
In the work, we consider two categories of neutron star matter: one contains just neutrons, protons, electrons and muons, which is denoted as np. And the other contains neutrons, protons, electrons, muons and hyperons, which is denoted as npH. We start our calculation with three parameter sets GM1, TM1 and NL3 using the relativistic mean field model involved U boson to describe neutron star matter. The equations of state are derived under satisfying constraints of the recent nuclear and astrophysics (see Fig. 1). In Fig. 1, the black, red, blue, magenta, olive and oyan colored lines correspond to the six cases of the U bosonic coupling constants $g_{u}$ = 0, 4, 7, 10, 20 and 40 GeV$^{-2}$ in npH matter, while the dot one corresponds to the case of $g_{u}$ = 0 GeV$^{-2}$ in np matter. We give three cases in Eq. (13) for the relativistic mean field theory: (i) the $\sigma$, $\omega$ and $\rho$ mesons are involved in the SU(6) spin-flavor symmetry based upon the naive quark model, which is denoted as SU6-no $\sigma^{*}$, $\phi$, (ii) the $\sigma$, $\omega$, $\rho$ including strange mesons $\sigma^{*}$ and $\phi$ are involved in the SU(6) spin-flavor symmetry, which is denoted as SU6, (iii) the $\sigma$, $\omega$, $\rho$, $\sigma^{*}$ and $\phi$ mesons are involved in the general SU(3) flavor symmetry, which is denoted as SU3. Between the same kind of line type in npH matter, the results of $g_{u}$ = 40 GeV$^{-2}$ and 0 GeV$^{-2}$ correspond to stiffer and softer equations of state, repectively. After the appearance of hyperons in Fig. 1, the equations of state in npH matter become obviously softer compared with these in np matter, and the gaps between the equations of state of np matter and npH matter are smaller and smaller with the values of $g_{u}$ rising, especially in the case of SU3.
\begin{table}[]
\centering
\caption{\label{tab:parameter}
Maximum masses $m_{max} = M_{max}/M_{\odot}$ and the corresponding radii R and Kepler frequencies $\nu_{k}$ of neutron stars for the U bosonic coupling constant $g_{u}$ = 0, 4, 7, 10, 20 and 40 GeV$^{-2}$, respectively.
}
{\begin{tabular}{@{}ccccccccccccc@{}}
\hline
\hline
&& \multicolumn{8}{c}{npH matter}   & \multicolumn{1}{c}{np matter}           \\
&&$g_{u}$ (GeV$^{-2}$)&        & 0     & 4      &7              & 10     & 20     &40             & 0               \\
\hline
GM1&SU6-no $\sigma^{*}$, $\phi$&$m_{max}$&  &1.802 &1.857 &1.896 &1.933 &2.051 &2.255       &2.361  \\
   &SU6                        &$m_{max}$&  &1.847 &1.897 &1.933 &1.968  &2.078 &2.271           &2.361\\
   &SU3                        &$m_{max}$&  &1.964 &1.996  &2.020 &2.043 &2.120  &2.266          &2.312\\
   &SU6-no $\sigma^{*}$, $\phi$&$R$ (km)&   &12.763 &12.965 &13.111  &13.262 &13.758 &14.647    &11.931  \\
   &SU6                        &$R$ (km)&   &12.558  &12.758 &12.910 &13.065 &13.552 &14.512     &11.931\\
   &SU3                        &$R$ (km)&   &11.127  &11.344 &11.501 &11.672  &12.208 &13.243    &11.530 \\
   &SU6-no $\sigma^{*}$, $\phi$&$\nu_{k}$ (Hz)& &931.079  &922.991 &917.104 &910.400 &887.373 &847.069  &1178.955\\
   &SU6                        &$\nu_{k}$ (Hz)& &965.623 &955.788 &947.901 &939.502  &913.746  &862.060  &1178.955\\
   &SU3                        &$\nu_{k}$ (Hz)& &1193.912   &1169.249  &1152.194  &1133.520  &1079.445  &987.844   &1228.212\\
\hline
TM1& SU6-no $\sigma^{*}$, $\phi$&$m_{max}$&  &1.667 &1.731 &1.776 &1.820 &1.957 &2.190           &2.179   \\
   & SU6                        &$m_{max}$&  &1.712 &1.771 &1.814 &1.856  &1.984 &2.207           &2.179 \\
   & SU3                        &$m_{max}$&  &1.732 &1.779  &1.812 &1.845 &1.949  &2.137          &2.094\\
   & SU6-no $\sigma^{*}$, $\phi$&$R$ (km)&   &13.482 &13.666 &13.812  &13.959 &14.412 &15.256    &12.375   \\
   & SU6                        &$R$ (km)&   &13.055  &13.263 &13.430 &13.602 &14.141 &15.079     &12.375 \\
   & SU3                        &$R$ (km)&   &11.019  &11.275 &11.461 &11.642  &12.237 &13.384    &11.721 \\
   & SU6-no $\sigma^{*}$, $\phi$&$\nu_{k}$ (Hz)& &824.845  &823.475 &821.058 &818.070 &808.466 &785.386  &1072.340\\
   & SU6                        &$\nu_{k}$ (Hz)& &877.143 &871.360 &865.427 &858.686  &837.671  &802.355  &1072.340\\
   & SU3                        &$\nu_{k}$ (Hz)& &1137.765   &1113.920  &1097.188  &1081.381  &1031.356  &944.047   &1140.338\\
\hline
NL3&SU6-no $\sigma^{*}$, $\phi$&$m_{max}$&  &1.983 &2.030 &2.064 &2.096 &2.200 &2.383           &2.774   \\
   &SU6                        &$m_{max}$&  &2.063 &2.102 &2.132 &2.160  &2.251 &2.418           &2.774 \\
   &SU3                        &$m_{max}$&  &2.291 &2.312  &2.327 &2.342 &2.392  &2.492          &2.732\\
   &SU6-no $\sigma^{*}$, $\phi$&$R$ (km)&   &13.425 &13.618 &13.752  &13.902 &14.355 &15.268    &13.295    \\
   &SU6                        &$R$ (km)&   &13.010  &13.255 &13.424 &13.580 &14.089 &14.997     &13.295\\
   &SU3                        &$R$ (km)&   &11.920  &12.070 &12.184 &12.301  &12.701 &13.482    &12.900 \\
   &SU6-no $\sigma^{*}$, $\phi$&$\nu_{k}$ (Hz)& &905.328  &896.214 &890.761 &883.326 &862.322 &818.337  &1086.499 \\
   &SU6                        &$\nu_{k}$ (Hz)& &967.838 &950.138 &938.688 &928.723  &897.237  &846.646  &1086.499\\
   &SU3                        &$\nu_{k}$ (Hz)& &1163.119   &1146.556  &1134.214  &1121.710  &1080.568  &1008.414  &1128.045\\
\hline
\hline
\end{tabular} \label{tap}}
\end{table}
After the numerical solution of the Tolman-Oppenheimer-Volkoff equations of hydrostatic equilibrium for the isolated cold neutron stars, combined with the recent constraints on the mass and radius inferred from PSRs J1614-2230, J0348+0432, J2215-5135, J0952-0607, J0740+6620 and J0030-0451, we obtained the mass-radius relation diagram (see Fig. 2). The maximum masses and the corresponding radii and Kepler frequencies for the three cases of each parameter set are listed in Table 2. As shown in Fig. 2 and Table 2, the maximum masses of neutron stars increase with the values of $g_{u}$ increasing in npH matter. It is clear that the maximum masses of npH matter for the GM1, TM1 and NL3 parameter sets using the different U bosonic coupling constants vary by big margins and lie in the scope of 1.667-2.492 $M_{\odot}$. The corresponding radii lie in the range 11.019 to 15.268 km. For the np matter, the maximum mass and the corresponding radius ranges at $g_{u}$ = 0 GeV$^{-2}$ are 2.094-2.774 $M_{\odot}$ and 11.530-13.295 km, respectively. In addition, the masses can simultaneously satisfy the constraints of the four massive pulsars other than PSR J0740+6620 in the three cases for the GM1 and TM1 parameter sets with $g_{u}$ = 40 GeV$^{-2}$ as well as for the NL3 parameter set with $g_{u}$ = 20 and 40 GeV$^{-2}$. The mass-radius relations can simultaneously satisfy the mass and radius constraints of the six pulsars only in the cases of SU3 for the NL3 parameter set with $g_{u}$ = 0, 4, 7 and 10 GeV$^{-2}$. In paticular, the radius at the canonical mass of a neutron star with 1.4 $M_{\odot}$ has a far greater influence than the radius at the maximum mass and lies in the scope of 12.957-17.825 km.

Fig. 3 displays the mass-frequency relation for the GM1, TM1 and NL3 parameter sets with the different U bosonic coupling constants, combined with the observational limits on the frequency inferred from the rapidly rotating pulsars PSR J1748-2446ad ($\nu$ = 716 Hz) and XTE J1739-285 ($\nu$ = 1122 Hz) (\citealt{Hessels+etal+2006, Kaaret+etal+2007}). However, Bult et al. observed XTE J1739-285 in 2020, they did not find the rotation frequency near 1122 Hz and concluded that it is unlikely to have the sub-millisecond rotation period 0.89 ms (\citealt{Bult+etal+2021}). We take $\nu$ = 1000 Hz as the threshold of the upper limit to limit our numerical results in Fig. 3.  As shown in Fig. 2 and Table 2, the Kepler frequencies at the maximum masses in npH matter decrease with the values of $g_{u}$ increasing and lie in the scope of 785.386 - 1193.912 Hz, which happen with a relatively large variation since the variations of the maximum masses obtained for all equation of states are greater. The Kepler frequencies at the maximum masses lie in the scope of 1072.340 - 1228.212 Hz with $g_{u}$ = 0 GeV$^{-2}$ in np matter. At the same value of the Kepler frequency, the growth of $g_{u}$ value makes the mass of neutron star increase. At the same value of the mass, the growth of $g_{u}$ value makes the Kepler frequency of neutron star decrease. In particular, the Kepler frequencies at the canonical mass are also much larger and lie in the scope of 496.962 - 802.806 Hz in npH matter. The corresponding values at the canonical mass lie in the scope of 666.514 - 770.347 Hz with $g_{u}$ = 0 GeV$^{-2}$ in np matter. In the absence of U bosons, for np matter, the Kepler frequencies can reach above 1000 Hz in the massive neutron stars in the three cases for the GM1, TM1 and NL3 parameter sets; for npH matter, the Kepler frequencies can also reach above 1000 Hz in the massive neutron stars only in the cases of SU3 for the GM1, TM1 and NL3 parameter sets. In the presence of U bosons, for npH matter, the Kepler frequencies can reach above 1000 Hz in the massive neutron stars only in the cases of SU3 for the GM1 and TM1 parameter sets with $g_{u}$ = 4, 7, 10 and 20 GeV$^{-2}$. They can also reach above 1000 Hz at $g_{u}$ = 40 GeV$^{-2}$ in the case of SU3 for the NL3 parameter set. The above results show that if the rapidly rotating pulsars with the frequencies above 716 Hz are detected in the future, which will help us to effectively constrain the specific range of the U bosonic coupling constants in npH matter.

Fig. 4 depicts the variation of the dimensionless tidal deformabilities with the masses of neutron stars for the GM1, TM1 and NL3 parameter sets with the different U bosonic coupling constants. The constraints on $\Lambda$ from the recent gravitational wave data are also depicted. The green and blue overlaid arrows represent the recent constraints on the $\Lambda_{1.4}$ from the GW170817 ($\Lambda_{1.4}=190^{+390}_{-120}$) and GW190814 ($\Lambda_{1.4}=616^{+273}_{-158}$) data (\citealt{Abbott+etal+2018, Abbott+etal+2020, Miyatsu+etal+2022}). The orange line is the upper limit on the dimensionless tidal deformability at the canonical mass from the initial GW170817 data, $\Lambda_{1.4}$  = 800 (\citealt{Abbott+etal+2017}). In Fig. 4, the dimensionless tidal deformabilities are monotonically decreasing with the masses of neutron stars. At the same value of the tidal deformability, the growth of $g_{u}$ value makes the mass of neutron star increase. At the same value of the mass, the growth of $g_{u}$ value makes the tidal deformability of neutron star increase.
In particular, the tidal deformabilities at the canonical mass are also much larger and lie in the scope of 461.273 - 3088.382 in npH matter, $\Lambda_{1.4}$ are consistent with the observed results from the GW170817, GW190814 and $\Lambda_{1.4} \leq$ 800 data only in the cases of SU3 for the GM1 and TM1 parameter sets with $g_{u}$ = 0 and 4 GeV$^{-2}$, $\Lambda_{1.4}$ in the cases of SU3 for the NL3 parameter set with $g_{u}$ = 0 and 4 GeV$^{-2}$ can only satisfy the constraints of the GW190814 and $\Lambda_{1.4} \leq$ 800 data. The corresponding values at the canonical mass lie in the scope of 469.244 - 1189.347 with $g_{u}$ = 0 GeV$^{-2}$ in np matter, $\Lambda_{1.4}$ can satisfy the constraints of the GW170817, GW190814 and $\Lambda_{1.4} \leq$ 800 data at the same time only in the cases of SU3 for the GM1 and TM1 parameter sets with $g_{u}$ = 0 GeV$^{-2}$, $\Lambda_{1.4}$ in the case of SU3 for the NL3 parameter set with $g_{u}$ = 0 GeV$^{-2}$ can only satisfy the constraint of the GW190814 and $\Lambda_{1.4} \leq$ 800 data. In the future, more and more the gravitational wave data related to neutron stars will be detected, which will be very meaningful for us to analyze the species of particles and the range of the U bosonic coupling constants in neutron stars.

\section{Conclusion}
In the work, we analysed the possible constraints on the range of the U bosonic coupling constants in neutron star containing hyperons which are combined with measurements and estimations of the mass and radius inferred from PSRs J1614-2230, J0348+0432, J2215-5135, J0952-0607, J0740+6620 and J0030-0451, the frequency inferred from the rapidly rotating pulsars PSR J1748-2446ad and XTE J1739-285 as well as gravitational wave data inferred from the events GW170817 and GW190814. The equations of state for neutron stars are obtained by the relativistic mean field model with the parameter sets of GM1, TM1 and NL3 under the SU(6) spin-flavor and SU(3) flavor symmetries. The results show that the growth of the U bosonic coupling constants $g_{u}$ makes the gaps between the equations of state of npH and np matter smaller. Especially when in the case of SU(3) flavor symmetry for the TM1 parameter set with $g_{u}$ = 40 GeV$^{-2}$, the two the equations of state for npH and np matter almost coincide. For npH matter, the masses and radii of neutron stars can simultaneously satisfy the constraint of the above six pulsars only in the cases of the SU(3) flavor symmetry for the NL3 parameter set with $g_{u}$ = 0, 4, 7 and 10 GeV$^{-2}$. The Kepler frequencies can reach above 1000 Hz only in the case of the SU(3) flavor symmetry for the GM1 and TM1 parameter sets without $g_{u}$ = 40 GeV$^{-2}$ and for the NL3 parameter set with $g_{u}$ = 0, 4, 7, 10, 20 and 40 GeV$^{-2}$. The tidal deformabilities can meet the constraints of the GW170817 and GW190814 data at the same time only in the case of the SU(3) flavor symmetry for the GM1 and TM1 parameter sets with $g_{u}$ = 0 and 4 GeV$^{-2}$.

Although our results depend on the adopted model and parameters, it is clear from the results that the U bosonic coupling constants based on the current observational data is very likely to be located in the range of 0 - 20 GeV$^{-2}$ from the three relations of the mass-radius, mass-frequency and mass-tidal deformability in neutron star containing hyperons. It might help us further constrain the equations of state of neutron stars containing hyperons. In addition, the numerical results of the baryonic and mesonic coupling constants under SU(3) flavor symmetry are more reasonable than those under SU(6) spin-flavor symmetry combined with the astronomical observations. We will continue to carry out in-depth research in the future. It is believed that the internal structure of neutron stars will finally be revealed with the continuous improvement of astronomical observations and nuclear physics data related to neutron stars.

\normalem
\begin{acknowledgements}
The authors sincerely thank Dr. J. L. Han from National Astronomical Observatory, Chinese Academy of Sciences for his constructive comments on the theoretical research of neutron stars.
\end{acknowledgements}

\section*{Data Availability}
The data are not publicly available due to the restrictions, e.g.,their containing information that could compromise the third-party rights and the privacy of research participants.

\bibliographystyle{raa}
\bibliography{bibtex}

\end{document}